# Spontaneous symmetry breaking induced unidirectional rotation of chain-grafted colloid in the active bath

Hui-shu Li[1], Chao Wang[1], Kang Chen[1*], Wen-de Tian[1*], and Yu-qiang Ma[1,2]

[1]Center for Soft Condensed Matter Physics & Interdisciplinary Research, College of Physics, Optoelectronics and Energy, Soochow University, Suzhou 215006, China.

[2]National Laboratory of Solid State Microstructures and Department of Physics, Nanjing University, Nanjing 210093, China

[*]Corresponding authors: kangchen@suda.edu.cn (K.C.) and tianwende@suda.edu.cn (W.d.T.)

Abstract

Exploiting the energy of randomly moving active agents such as bacteria is a fascinating way to power a microdevice. Here we show, by simulations, that a chain-grafted disk-like colloid can rotate unidirectionally when immersed in a thin film of active particle suspension. The spontaneous symmetry breaking of chain configurations is the origin of the unidirectional rotation. Long persistence time, large propelling force and/or small rotating friction are keys to keeping the broken symmetry and realizing the rotation. In the rotating state, we find very simple linear relations, e.g. between mean angular speed and propelling force. The time-evolving asymmetry of chain configurations reveals that there are two types of non-rotating state. Our findings provide new insights into the phenomena of spontaneous symmetry breaking in active systems with flexible objects and also open the way to conceive new soft/deformable microdevices.

Self-propelling motions are ubiquitous in biological world, ranging from the molecular-level transportation of motor proteins along the microtubules, to the swimming of bacteria on the micrometer scale [1-4]. Such non-thermal motion leads to very rich out-of-equilibrium phenomena, e.g. giant fluctuations [5,6], spontaneous phase separation [7] and pattern formation [8]. Apart from understanding the foundations of physics in these phenomena, an intriguing topic is to design microdevices or micromotors that can rectify the random motions and convert the energy into mechanical work [9]. Rectification of the random motions requires breaking of both the time and spatial symmetries [10-17]. The self-propelling motions of active agents break the time symmetry by irreversible energy consumption. The breaking of spatial symmetry is then typically realized by introducing shape-asymmetric objects, e.g. microscopic gears with asymmetric teeth [11,18-20].

Is the built-in shape-asymmetry the prerequisite for such microdevices? Instead of rigid body, is it possible to make a soft/deformable microdevice? Softness/deformability may bring the advantages of damage resistance, durability, adaptability and so on. Recently, the behaviors of chain-like object [21-30] or deformable boundary [31-33] in the active system have received much attention. For a passive chain in contact with the active bath [21,23,31,32], additional non-thermal fluctuations on chain beads due to collisions with the self-propelling particles cause the anomalous deformation of the chain. Meanwhile, the shape of the chain (local curvature) influences the trapping of the motile particles in its vicinity [34,35]. How about transplant the softness/deformability to the microdevice by introducing the chain-like structures?

In this paper, we study by numerical simulations a disk-like colloid with chains grafted on its side immersed in a thin film of active particle suspension. Surprisingly, we find, under certain conditions, the grafted chains, in cooperation with the collective motion of trapped active particles, spontaneously form longstanding asymmetric configurations, which leads to unidirectional rotation of the colloid. In contrast to the rotary microgears in the bacterial bath that the broken spatial symmetry (asymmetric teeth) is fixed in advance, no built-in spatial asymmetry is introduced on purpose in our system. It's of fundamental interest to learn how such asymmetry comes into being and is then maintained.

In our model, active bath particles and passive beads of the grafted chains are considered as small disks of mass $m$ and size $\sigma$. The colloid is treated as a rigid circle of radius $R = 5\sigma$ with the rim mimicked by closely connected small disks of size $\sigma$. $n_c = 15$ chains, each consisting of $N$ beads, are uniformly grafted on its surface. All disks interact with repulsive Weeks-Chandler-Andersen (WCA) potential, $U_{WCA}(r) = 4\varepsilon \left[ (\sigma/r)^{12} - (\sigma/r)^6 \right] + \varepsilon$ with a cutoff at $r = 2^{1/6}\sigma$ beyond with $U_{WCA} = 0$. $m$, $\sigma$ and $\varepsilon$ are used as scaling units in the simulation. Additional bonded harmonic potential between chain beads is $U_b = k(r - r_0)^2$, where $k=1000$ is the spring constant and $r_0 = 0.98$ the equilibrium bond length. The motion of the active bath particles and chain beads is described by the Langevin equations,

$$m\ddot{\vec{r}}_i = -\nabla_i U_i - \zeta \dot{\vec{r}}_i + F\hat{\mu}_i(t) + \sqrt{2\zeta k_B T}\vec{\eta}_i(t) \qquad (1)$$

$$\dot{\theta}_i = \sqrt{2D_r}\xi_i(t) \qquad (2)$$

Eq. (1) governs the translational motion. For the chain beads, the propelling force $F = 0$ and only Eq. (1) is required. Eq. (2) depicts the coupled rotational kinetics of the driving direction $\hat{\mu}_i = (\cos\theta_i, \sin\theta_i)$ for active particles. $\zeta$ and $D_r$ are translational friction coefficient and rotational diffusion rate, respectively. $\vec{\eta}_i(t)$ and $\xi_i(t)$ are unit-variance

Gaussian white noises. We fix the center of the colloid to focus on its rotational motion, which is described by

$$I\dot{\omega} = -\zeta_R \omega + M \qquad (3)$$

$I = 25$ is the rotational inertia, $\zeta_R$ the rotational friction coefficient and $M$ the total torque. Unless otherwise stated, we set $\zeta = 10$ and $\zeta_R = 100$ which are large that the motions are essentially overdamped. For simplicity, we ignore the translational thermal fluctuation in Eq. (1). The simulation box size is 130x130 and the area fraction of active particles is as low as 0.05.

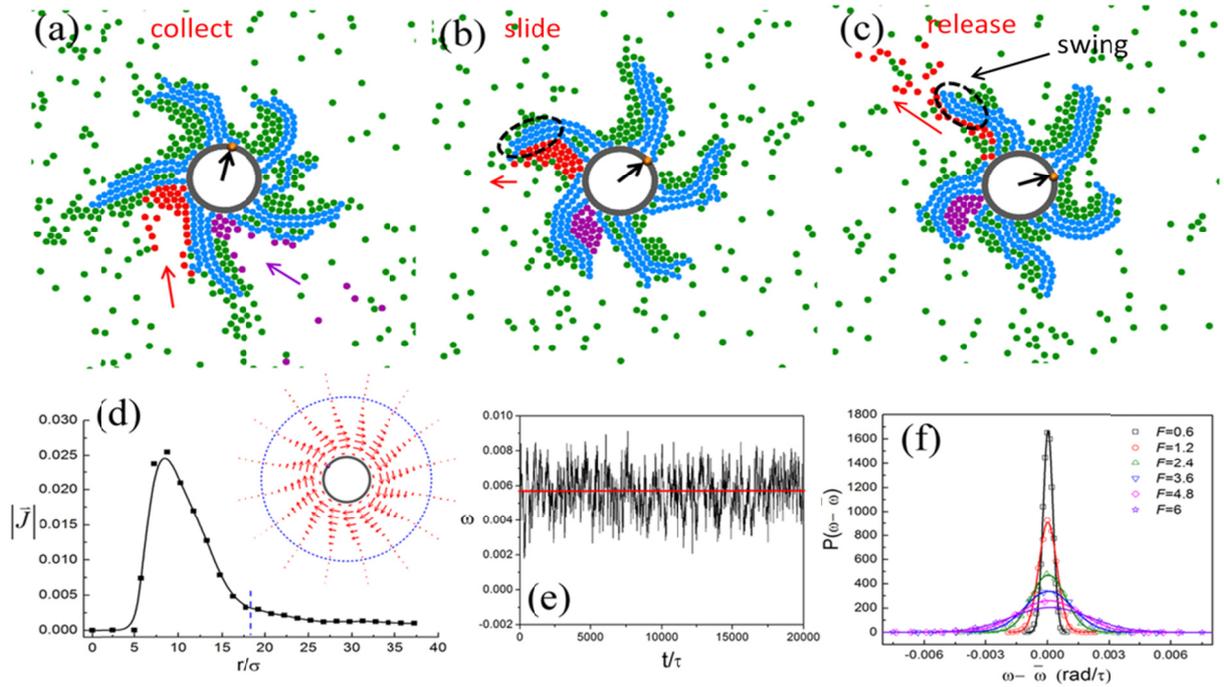

**Fig. 1**: (a)-(c) Snapshots at sequential times. The solid circles represent active particles (green, red and purple) and chain beads (light blue), respectively. The large hollow grey circle represents the colloid and the black arrow at the center shows the clockwise rotation. The circles in red and purple show the updating process of trapped particles: collection, sliding toward the free ends of chains and release. The free ends circled with dashed lines in (b) and (c) exemplify their swinging, accompanying the particle release. (d) The magnitude of the averaged flow of active particles as a function of distance from the center of the colloid. The picture shows the pattern of the averaged flow. The length of arrow is proportional to magnitude of flow. The blue dashed lines mark roughly the top of the brush. (e) A representative time sequence of instantaneous angular velocity for clockwise (positive) rotation, where the horizontal line is the mean angular velocity, $\bar{\omega}$. (f) The probability distribution of $\omega$ around $\bar{\omega}$ for different propelling forces. The solid lines are fittings by Gaussian distribution. The parameters $F=3.6$ for (a) to (e), $N$=15 and $D_r=3.6\times10^{-4}$ for (a) to (f).

We find that, in the case of $F=3.6$ and $D_r=3.6\times10^{-4}$, the colloid rotates unidirectionally, driven by the active particles trapped around the grafted chains (see Fig. 1(a)-(c) and supplemental movie1 [36]). Apparently, the grafted chains assemble into loose and unstable bundles, so that most active particles in the "brush regime" are trapped in the interval between bundles. Intriguingly, asymmetric shape of these chain bundles is formed spontaneously. And once the bias of the shape appears, it is maintained, in spite of the significant shape fluctuation. The bias and hence the direction of rotation of the colloid is random, determined by initial condition and fluctuation. Active particles tend to accumulate at corners or the places of large curvature on the boundary or surface of an obstacle [34,35]. Therefore, the particles entering the brush regime are collected depending on their moving directions and mostly trapped on the concave side of the bundles. The trapped particles are updated frequently. Along with the rotation of the colloid, they slide toward the free ends of grafted chains and then are released, often accompanied by the swinging of the free ends.

The start of the spontaneous symmetry breaking or unidirectional rotation is shown in the supplemental movie2 [36]. In the beginning, the active particles aggregate densely (as densely as in the supplemental movie5 or movie6 [36]) in the brush regime which induces collective motion of both the active particles and chain beads. Randomly, the push in one direction dominates and then the colloid starts to rotate and the symmetry of chain configurations breaks. After rotation, the initial trapped particles are updated and the new trapped particles are collected biasedly depending on the asymmetric configuration of chains. As a consequence, the number of trapped particles decreases after rotation. This is a kind of positive-feedback process which eventually causes the longstanding broken symmetry of the brush and the unidirectional rotation of the colloid.

The picture in Fig. 1(d) shows the averaged flow pattern of active particles around the colloid, $\vec{J} = \langle \overline{\rho_a \vec{v}_a} \rangle$ where $\rho_a$ is the number density of active particles and $\vec{v}_a$ the velocity. The overbar and angular bracket represent averaging over time and six independent runs, respectively. Unidirectional circular current is formed in the brush regime (within the blue dashed line) and peaked at the place closer to the colloidal surface than to the brush top. Such circular current causes the unidirectional rotation of the colloid. The angular velocity of the rotation $\omega$ fluctuates around the mean value (Fig. 1(e)) and the probability distribution is roughly left-right symmetric (Fig. 1(f)), which can be well fit by Gaussian distribution. As the propelling force increases, the distribution (fluctuation) becomes broader (larger). In the macroscopic experiment of robot-driven rotary gear [12], such distribution also becomes more and more left-right symmetric as the number of robots increases.

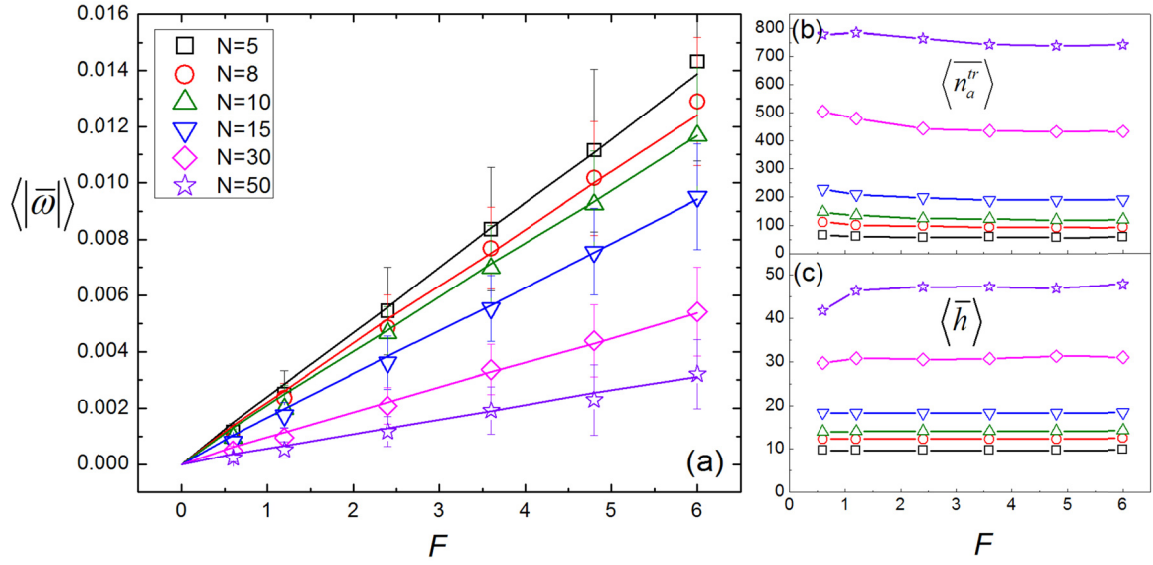

**Fig. 2**: The mean angular speed (a), number of trapped active particles (b) and "middle" height of brush (c) as functions of propelling force for various chain lengths. The solid lines in (a) are fittings by Eq. (4). The parameter $D_r = 3.6 \times 10^{-4}$.

The dependence of the mean angular speed on the propelling force and the length of grafted chains is shown in Fig. 2(a). Interestingly, very good linear relation is found between $\langle |\bar{\omega}| \rangle$ (|…| represents absolute value) and $F$ for all chain lengths. This linear relation can be understood by the following analysis. The trapping of active particles in the brush regime satisfies the strong confinement condition since the persistence length of active particles, $F/(\zeta D_r)$, is much larger than the interval between bundles. Then, approximately, the active pressure on the brush $P_a \propto \langle \overline{\rho_a^{tr}} \rangle F \Delta$ [35,37,38], where $\langle \overline{\rho_a^{tr}} \rangle$ is the mean density of trapped active particles in the brush regime and $\Delta$ the characteristic width of the interval between bundles. This expression corresponds to each trapped particle exerts $F$ (the propelling force) on the brush. Concerning the total force on the brush along the tangential direction (the center of the colloid is treated as the origin) due to active motion of particles, we have $F_a^\perp \approx \alpha \langle \overline{n_a^{tr}} \rangle F$, where $\langle \overline{n_a^{tr}} \rangle$ is the mean number of active particles trapped in the brush regime and $\alpha$ is a prefactor accounting for the average tangential component of the propelling force. Along with the rotation, there are viscous forces on the active particles and chain beads. Approximately, the viscous force in the tangential direction $F_v^\perp \approx -\left(\langle \overline{n_a^{tr}} \rangle + n_c N\right) \zeta \langle |\bar{\omega}| \rangle \langle \bar{h} \rangle$, where $h \equiv \sum_i^{n_c} \sum_j^N h_{ij} / (n_c N)$ is the mean distance of chain beads to the center of the colloid, i.e. $\langle \bar{h} \rangle$ is the mean "middle" height of the brush. The torque on the colloid due to the motion of trapped particles and chain beads is then given by $M \approx \left(F_a^\perp + F_v^\perp\right)\langle \bar{h} \rangle$. In the overdamped limit, this torque is balanced by the viscous torque on the colloid due to rotation $M' = -\zeta_R \langle |\bar{\omega}| \rangle$. Therefore, we obtain

$$\langle |\bar{\omega}| \rangle \approx \frac{\alpha \langle \overline{n_a^{tr}} \rangle \langle \bar{h} \rangle}{\zeta_R + \left(\langle \overline{n_a^{tr}} \rangle + n_c N\right) \zeta \langle \bar{h} \rangle^2} F \qquad (4)$$

Calculation shows that the quantities $\langle \overline{n_a^{tr}} \rangle$ and $\langle \bar{h} \rangle$ are nearly independent of $F$ (Fig. 2(b) and (c)), i.e. for a certain chain length $N$, the factor in front of $F$ on the right-hand side of Eq. (4) is nearly a constant, or $\langle |\bar{\omega}| \rangle$ is proportional to $F$ with the factor as the slope. We use $\alpha$ as the only fitting parameter and the solid lines in Fig. 2(a) are fittings by Eq. (4) which agrees very well with simulation data. The parameter $\alpha$ varies in a narrow range around 0.5 to 0.63 for different $N$'s. From the data in Fig. 2(b) and (c), we obtain approximate expressions at large $F$: $\langle \overline{n_a^{tr}} \rangle \approx 15.4N - 29.5$ and $\langle \bar{h} \rangle \approx 0.830N + 5.74$. For large $N$, the numerator of the factor in Eq. (4) $\propto N^2$, while the denominator $\propto N^3$. This explains why the slope in Fig. 2(a) decreases with increasing $N$. Intuitively, the slope should reach a maximum value at a certain small $N$, denoting the most efficient rotator. This optimized $N$ is the resultant of two competing factors: the requirement of gathering enough active particles to provide the driving torque versus the dragging torque due to the viscous forces on both the active particles and chain beads. Here, however, we do not obtain such optimized $N$. Because when $N<5$, the rotation becomes unstable, i.e., the rotation may keep in one direction for a significant period then reverse due to occasional event (fluctuation).

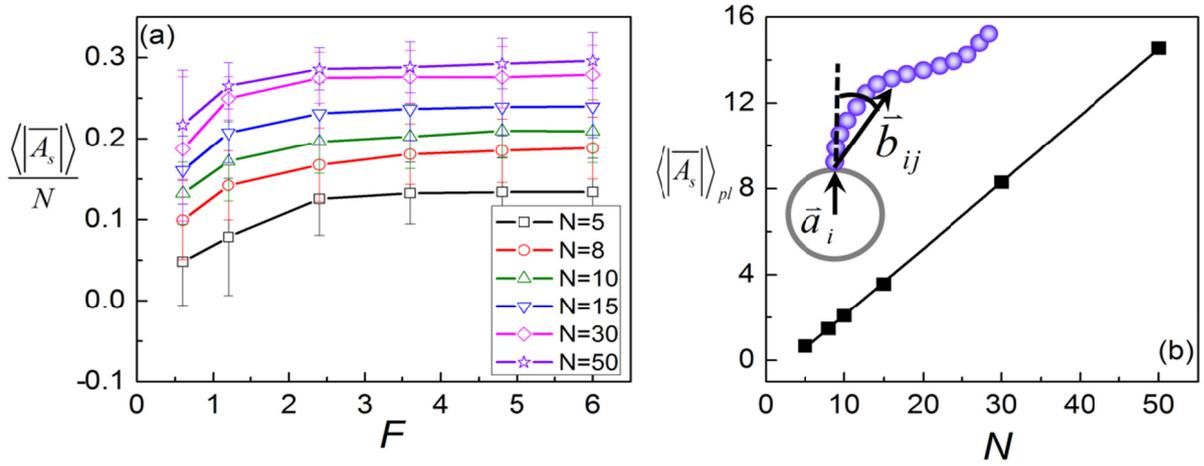

**Fig. 3**: (a) The mean degree of asymmetry of the chain configurations normalized by chain length as a function of propelling force for various chain lengths. (b) The plateau value of the mean degree of asymmetry as a function of chain length. The solid line gives the expression $0.31N-1.0$. The inset picture is a schematic showing the calculation of $A_s$. The parameter $D_r = 3.6 \times 10^{-4}$.

Spontaneous symmetry breaking of the configuration of grafted chains is the prerequisite for the unidirectional rotation of the colloid. To quantify the degree of asymmetry, we define $A_s = \sum_{i}^{n_c} \sum_{j}^{N} \vec{a}_i \times \vec{b}_{ij} \big/ (n_c N R)$, where $\vec{a}_i$ is the vector from the center of the colloid to the grafted point of the $i$th chain and $\vec{b}_{ij}$ the vector from the grafted point to the $j$th bead as shown in the schematic of Fig. 3(b). We find that $\langle |\overline{A_s}| \rangle$ also weakly depends on $F$ and notable increment is found only at small $F$, which is compatible with the $F$-variation of $\langle \overline{n_a^{tr}} \rangle$ and $\langle \overline{h} \rangle$ (Fig. 3(a)). At large $F$, $\langle |\overline{A_s}| \rangle$ reaches a plateau. This plateau value versus the brush chain length follows a perfect linear relation, $\langle |\overline{A_s}| \rangle_{pl} \approx 0.31N - 1.0$ (Fig. 3(b)).

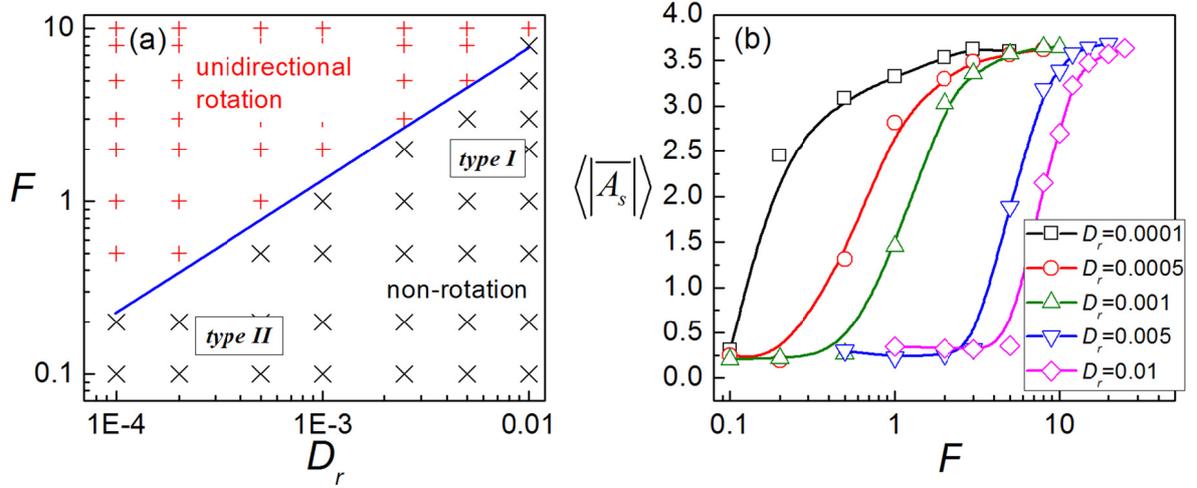

**Fig. 4**: (a) phase diagram in $F$-$D_r$ space. (b) The mean degree of asymmetry as a function of propelling force for various $D_r$. The parameter $N=15$.

To explore the requisites for generating the unidirectional rotation of the hairy colloid, we calculate the phase diagram for $N=15$ in Fig. 4(a) spanned by two important parameters $F$ (amplitude of propelling force, equivalent to swimming speed at a certain $\zeta$) and $D_r$ (inverse proportional to persistence time), which characterize the propelling property of the active agents. Not surprisingly, unidirectional rotation (rotating state) happens at large $F$ and small $D_r$, i.e. in the situation of strong activity. The shape of the corresponding $F$-dependent $\langle |\overline{A_s}| \rangle$ that crosses the phase boundary is sigmoidal (Fig.4 (b)), depicting the drastic shift of the brush asymmetry in the non-rotation to rotation transition. In the rotating state, $\langle |\overline{A_s}| \rangle$'s for different $D_r$'s all saturate around the plateau value $\langle |\overline{A_s}| \rangle_{pl} \approx 3.5$ at large force, implying similar asymmetric shape of the brush in the plateau regime. The independence of $\langle |\overline{A_s}| \rangle_{pl}$ on $D_r$ also suggests the linear relation $\langle |\overline{A_s}| \rangle_{pl} \approx 0.31N - 1.0$ is universal.

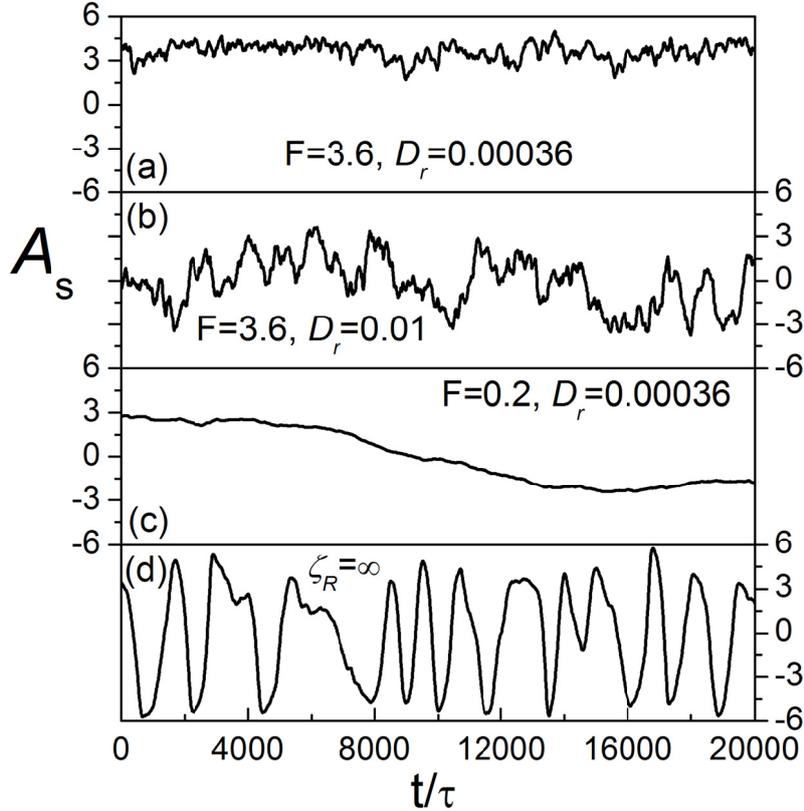

**Fig. 5**: Representative time evolution of $A_s$ in the rotating (a), non-rotating states (b)-(c) and the special case of infinite viscous drag on the rotation of the colloid (d).

We compare the representative time evolution of $A_s$ in the rotating and non-rotating states in Fig. 5. $A_s$ fluctuates moderately around a nonzero mean value (i.e. there is longstanding broken symmetry) in the rotating state ($F$=3.6 and $D_r = 3.6 \times 10^{-4}$) in Fig. 5(a), while it fluctuates strongly around zero (i.e. there is not longstanding broken symmetry) in the non-rotating state when solely increasing the rotational diffusion coefficient to be, for example, $D_r = 0.01$ (Fig. 5(b)). Larger $D_r$ means smaller persistence time or more frequent reorientation of the active particles. Therefore when the trapped particles reorient before escaping the brush regime, they are not able to generate longstanding broken symmetry of the brush in one direction, and hence the unidirectional rotation (see supplemental movie3 [36]). Since the trapping time decreases with increasing $\omega$ or $F$, the crossover $D_r$ (at the phase boundary) increases with $F$ as in the phase diagram. The $A_s$ curve in Fig. 5(b) represents one typical type of non-rotating state (denoted as type I) which is at the upper-right part of the phase diagram. On the other hand, if we decrease the propelling force alone to be, for example, $F=0.2$, $A_s$ varies slowly with time instead of strongly fluctuating (see Fig. 5(c) and supplemental movie4 [36]). Quasi-longstanding broken symmetry of brush is indeed formed, however the small $F$ hence the small torque does not produce notable unidirectional rotation of the colloid, and meanwhile the motion of chains eventually reverses the direction of asymmetry. This kind of $A_s(t)$ characterizes another type of non-rotating state (type II) which is at the lower-left part of the phase diagram. Notable reversing process of the asymmetry and also the rotating direction is found when we increase the viscous-drag coefficient $\zeta_R$ for the rotation of the colloid. In the case of $\zeta_R = 5000$ the rotation of the colloid is not quick enough to catch up with the motion of chains and leads to the reversion (see supplemental movie5 [36]). In the extreme condition that the colloid is not allowed to rotate $A_s$ oscillates more regularly and with larger amplitude (Fig. 5(d)). Correspondingly, the grafted chains vibrate like the beating of cilia (see supplemental movie6 [36]).

In summary, we have shown that disk-like colloid with chains grafted on its side can rotate unidirectionally when immersed in a thin film of active particle suspension. The fundamental reason is the spontaneous symmetry breaking of chain configurations and the maintenance of this broken symmetry. Hydrodynamic interactions are omitted in our "dry" model, but we believe the phenomena and main conclusions are qualitatively valid in both "dry" and "wet" systems [18]. Furthermore, our "dry" model is realizable in macroscopic granular or robotic systems. Our work puts forward a new design concept of making soft/deformable microdevices which may have advantages of damage resistance, durability, adaptability etc. and also provide new insights into the spontaneous symmetry breaking phenomena in active systems with the presence of flexible objects.

This work is supported by the National Natural Science Foundation of China (NSFC) Nos. 21374073 (K.C.), 21574096 (K.C.) and 21474074 (W.-d.T.).